 \documentclass[final,1p,times]{elsarticle}

\begin{document}

\begin{frontmatter}

\title{Search for a diffuse flux of high-energy  $\nu_\mu$  with the ANTARES neutrino telescope}
\begin{abstract}
A search for a diffuse flux of astrophysical muon neutrinos, using data collected by the ANTARES neutrino telescope  is presented. A $(0.83\times 2\pi)$ sr sky was monitored for a total of 334 days of equivalent live time.
The searched signal corresponds  to  an excess of events,  produced  by astrophysical sources, over the expected atmospheric neutrino background.
The observed number of events is found compatible with the background expectation. Assuming an  $E^{-2}$  flux spectrum, a 90\% c.l. upper limit on the diffuse $\nu_\mu$ flux of $E^2\Phi_{90\%}  =   5.3 \times 10^{-8}   \  \mathrm{GeV\ cm^{-2}\ s^{-1}\ sr^{-1}} $ in the energy range 20 TeV - 2.5 PeV is obtained. 
Other signal models with different energy spectra are also tested and some rejected.  
\end{abstract}

\author[IFIC]{J.A. Aguilar}
\author[CPPM]{I. Al Samarai}
\author[Colmar]{A. Albert}
\author[Barcelona]{M.~Andr\'e}
\author[Genova]{M. Anghinolfi}
\author[Erlangen]{G. Anton}
\author[IRFU/SEDI]{S. Anvar}
\author[UPV]{M. Ardid}
\author[NIKHEF]{A.C. Assis Jesus}
\author[NIKHEF]{T.~Astraatmadja\fnref{tag:1}}
% ...
\newpage
\address[IFIC]{\scriptsize{IFIC - Instituto de F\'isica Corpuscular, Edificios Investigaci\'on de Paterna, CSIC - Universitat de Val\`encia, Apdo. de Correos 22085, 46071 Valencia, Spain}}
\address[CPPM]{\scriptsize{CPPM, Aix-Marseille Universit\'e, CNRS/IN2P3, Marseille, France}}
\address[Colmar]{\scriptsize{GRPHE - Institut universitaire de technologie de Colmar, 34 rue du Grillenbreit BP 50568 - 68008 Colmar, France }}
\address[Barcelona]{\scriptsize{Technical University of Catalonia,Laboratory of Applied Bioacoustics,Rambla Exposici\'o,08800 Vilanova i la Geltr\'u,Barcelona, Spain}}
\address[Genova]{\scriptsize{INFN - Sezione di Genova, Via Dodecaneso 33, 16146 Genova, Italy}}
\address[Erlangen]{\scriptsize{Friedrich-Alexander-Universit\"{a}t Erlangen-N\"{u}rnberg, Erlangen Centre for Astroparticle Physics, Erwin-Rommel-Str. 1, 91058 Erlangen, Germany}}
\address[IRFU/SEDI]{\scriptsize{Direction des Sciences de la Mati\`ere - Institut de recherche sur les lois fondamentales de l'Univers - Service d'Electronique des D\'etecteurs et d'Informatique, CEA Saclay, 91191 Gif-sur-Yvette Cedex, France}}
\address[UPV]{\scriptsize{Institut d'Investigaci\'o per a la Gesti\'o Integrada de Zones Costaneres (IGIC) - Universitat Polit\`ecnica de Val\`encia. C/  Paranimf 1. , 46730 Gandia, Spain.}}
\address[NIKHEF]{\scriptsize{Nikhef, Science Park,  Amsterdam, The Netherlands}}
% ...
\address[IRFU/SPP]{\scriptsize{Direction des Sciences de la Mati\`ere - Institut de recherche sur les lois fondamentales de l'Univers - Service de Physique des Particules, CEA Saclay, 91191 Gif-sur-Yvette Cedex, France}}
% ...
\address[Genova-UNI]{\scriptsize{Dipartimento di Fisica dell'Universit\`a, Via Dodecaneso 33, 16146 Genova, Italy}}

\fntext[tag:1]{\scriptsize{Also at University of Leiden, the Netherlands}}
\fntext[tag:2]{\scriptsize{On leave at DESY, Platanenallee 6, D-15738 Zeuthen, Germany}}
\cortext[ca]{Corresponding authors. Contact:spurio@bo.infn.it,phone:+39-051-2095248}

\end{frontmatter}


\section{Introduction} 

This letter presents a search for a diffuse flux of high energy muon neutrinos from astrophysical sources with the ANTARES neutrino telescope. The construction of the
deep sea ANTARES detector was completed in May 2008 with the connection of its twelfth detector line. The telescope is located 42 km off the southern coast of France, near Toulon, at a maximum depth of 2475 m. 

The prediction of the diffuse neutrino flux from unresolved astrophysical sources is based on cosmic ray (CR) and $\gamma$-ray observations.
Both electrons ({\it leptonic models}) \cite{rnc,ropp} and protons or nuclei ({\it hadronic models}) \cite{adro} can be accelerated in astrophysical processes.  
In the framework of hadronic models 
the energy escaping from the sources is distributed between CRs, $\gamma$-rays and neutrinos. 
Upper bounds for the neutrino diffuse flux are derived from the observation of the diffuse fluxes of $\gamma$-rays and  ultra high energy CRs taking into account the production kinematics, the opacity of the source to neutrons and the effect of propagation in the Universe.
There are two relevant predictions:

\noindent \textit{-- The Waxman-Bahcall (W\&B) upper bound} \cite{wb} uses the CR observations at $E_{CR}\sim 10^{19}$ eV ($E_{CR}^2 \Phi_{CR} \sim 10^{-8}$ GeV cm$^{-2}$s$^{-1}$sr$^{-1}$) to constrain the diffuse flux per neutrino flavour (here and in the following the symbol $\nu$   represents the sum of  $\nu_\mu$ plus $\overline \nu_\mu$): 
\begin{equation}
E^2_\nu \Phi_\nu < 4.5/2 \times 10^{-8} \
\textrm{GeV}\  \textrm{cm}^{-2} \textrm{sr}^{-1}\textrm{s}^{-1}
\label{eq:wb}
\end{equation}
(the factor 1/2 is added to take into account neutrino oscillations). This value represents a benchmark flux for neutrino telescopes. 

\noindent \textit{-- The Mannheim-Protheroe-Rachen (MPR) upper bound} \cite{mpr} is derived using as  constraints the observed CR fluxes over the range from 10$^5$ to 10$^9$  GeV and $\gamma$-ray diffuse fluxes.
In the case of sources \textit{opaque} to neutrons, the limit is $E^2_\nu \Phi_\nu < 2 \times 10^{-6}\
(\textrm{ GeV}\  \textrm{cm}^{-2} \textrm{sr}^{-1}\textrm{s}^{-1})$; in the case of sources \textit{transparent} to neutrons, the limit decreases from the value for \textit{opaque}   sources at $E_\nu \sim  10^6$ GeV to the value of Eq. \ref{eq:wb} at $E_\nu \sim  10^9$ GeV.
  
The detection of high energy cosmic neutrinos is not background free. Showers induced by interactions of CRs with the Earth's atmosphere give rise to {\it atmospheric muons} and {\it atmospheric neutrinos}. Atmospheric neutrinos that have traversed the Earth and have been detected in the neutrino telescope, are an irreducible background for the study of cosmic neutrinos. As the spectrum of cosmic neutrinos is expected to be harder ($\propto E_\nu^{-2}$) than that of atmospheric neutrinos, a way to distinguish the $\nu_\mu$  cosmic diffuse flux is to search for an excess of high energy events in the measured energy spectrum.

The relevant characteristics  of the ANTARES detector are presented in Sec. 2.  
The rejection of the atmospheric muon background and an estimator of the muon energy are discussed in Sec. 3.  This estimator is used to discriminate high energy neutrino candidates from the bulk of lower energy atmospheric   neutrinos.  The results are  presented and discussed in Sec. 4 and Sec. 5.

\section{$\nu_\mu$ reconstruction in the ANTARES detector} 

The ANTARES detector is a three-dimensional array of photomultiplier tubes (PMTs) distributed along twelve lines \cite{anta_line1}. 
Each line comprises 25 storeys, spaced vertically by 14.5 m, with each storey containing three optical modules (OMs) \cite{anta_om} and a local control module for the corresponding electronics \cite{antaresDAQ}. The OMs (885 in total) are arranged with the axes of the PMTs oriented 45$^\circ$ below the horizontal.  
The lines are anchored on the seabed at distances of about 70 m from each other and tensioned by a buoy at the top of each line.

Muon neutrinos are detected via charged current interactions: $\nu_\mu + N \rightarrow \mu+X$. The challenge of measuring muon neutrinos consists of reconstructing the trajectory using the arrival times and the amplitudes of the Cherenkov light signal detected by the OMs, and of estimating the energy. 
The track reconstruction algorithm \cite{aart_icrc09} is based on a likelihood fit that uses a detailed parametrization of the probability density function for the photon arrival times taking into account the delayed photons. The output is: the track position and direction; the information on the number of hits ($N_{hit}$) used for the reconstruction; a quality parameter $\Lambda$.  $\Lambda$ is determined from the likelihood and the number of compatible solutions found by the algorithm and can be used to reject badly reconstructed events. Without any cut on $\Lambda$, the fraction of atmospheric muon events that are reconstructed as upward-going is $\sim$ 2\% (see Table \ref{tab:prelim}). The appropriate value of the $\Lambda$ variable cut for this analysis is discussed in Sec. 3.
Monte Carlo (MC) simulations show that the ANTARES detector achieves   a median angular resolution for muon neutrinos better than 0.3$^\circ$ for $E_\nu>$ 10 TeV.

Muon energy losses are due to several processes \cite{murange} and can be  parametrized as:
\begin{equation}\label{muelos}
dE_\mu/dx= \alpha (E_\mu) + \beta (E_\mu) \cdot E_\mu   \ ,
\end{equation}
\noindent where $\alpha(E_\mu)$ is an almost constant term that accounts for ionisation, and $\beta(E_\mu)$ takes into account the radiative losses that dominate for $E_\mu>$0.5 TeV. 
Particles above the Cherenkov  threshold produce a coherent radiation emitted in a Cherenkov cone with a characteristic angle 
$\theta_C\simeq 43^\circ$ in water.  
Photons emitted at the Cherenkov angle, arriving at the OMs without being scattered, are referred to as \textit{direct photons}. The differences between the calculated and the measured arrival time (time residuals) of direct photons follow a nearly Gaussian distribution of few ns width, due to the chromatic dispersion in the sea water and to the transit time spread of the PMTs.  

For high muon energies ($E_\mu> 1$ TeV), the contribution of the energy losses due to radiative processes increases  linearly with the muon energy and the resulting electromagnetic showers produce additional light. 

Scattered Cherenkov radiation or photons originating from secondary electromagnetic showers arriving on the OMs (denoted from now on as \textit{delayed photons}), are delayed with respect to the \textit{direct photons}, with arrival time differences up to hundreds of ns \cite{chiarusi}. 
As a consequence, the percentage of delayed photons with respect to direct photons increases with the muon energy.

The PMT signal is processed by two ASIC chips (the Analogue Ring Sampler, ARS \cite{ars_paper}) which digitize the time and the amplitude of the signal (the $hit$). They are operated in a token ring scheme. If the signal crosses a preset threshold, typically 0.3 photo-electrons, the first ARS integrates the pulse within a window of 25 ns and then hands over to the second chip with a dead time of 15 ns. 
If triggered, the second chip provides a second hit with a further integration window of 25 ns. After digitization, each chip has a dead time of typically 250 ns. After this dead time, a third and fourth hit can also be present.

\subsection{The Monte Carlo  simulations} \label{MC}
The simulation chain \cite{brunner,mar5line} comprises the generation of Cherenkov light, the inclusion of the optical background  caused by bioluminescence and radioactive isotopes present in sea water, and the    digitization of the PMT signals. Upgoing muon neutrinos and downgoing atmospheric muons have been simulated and stored in the same format used for data. 

\noindent \textbf{Signal and atmospheric neutrinos.}
MC muon neutrino events have been generated in the  energy range  $10 \leq E_\nu \leq 10^8$  GeV and zenith angle between $0^\circ \leq \theta \leq 90^\circ $ (upgoing events). 
The same MC sample can be differently weighted to reproduce the ``conventional'' atmospheric neutrinos from charged meson decay (Bartol) \cite{bartol} ($\Phi_\nu\propto E_\nu^{-3.7}$ at high energies), the  ``prompt'' neutrinos and the theoretical astrophysical signal ($\Phi_\nu\propto E_\nu^{-2}$). 
A test spectrum:
\begin{equation}
E_\nu^2\ \Phi_{\nu} =  1.0 \times 10^{-7} \  \mathrm{GeV\ cm^{-2}\ s^{-1}\ sr^{-1}}.
\label{test_limit}
\end{equation}
is used to simulate the {diffuse flux signal}. The normalization of this test flux  is irrelevant when defining cuts,   optimizing procedures,  and calculating  the sensitivity.

Above 10 TeV,  the semi-leptonic decay of short-lived charmed particles $D \rightarrow K + \mu + \nu_\mu$ becomes a significant source of atmospheric ``prompt leptons''.
The lack of precise information on high-energy charm production in hadron-nucleus collisions leads to a great uncertainty (up to four orders of magnitude) in the estimate of the leptonic flux above 100 TeV.  The models considered in \cite{prompt_Costa} were used, in particular the Recombination Quark Parton Model (RQPM) which gives the largest prompt contribution. 

\noindent \textbf{Atmospheric muons.}
Atmospheric muons reconstructed as upgoing are the main background for a neutrino signal and their rejection is a crucial point in this analysis. Atmospheric muon samples have been simulated with the MUPAGE package \cite{mupage}. In addition to one month of equivalent live time with a total energy $E_T\geq$ 1 GeV \cite{antares_low}, a dedicated one year of equivalent live time with $E_T\geq$ 1 TeV and multiplicity $m = 1 \div 1000$ was generated. The total energy $E_T$ is the sum of the energy of the individual muons in the bundle. 
Triggered ANTARES  events mainly consist of multiple muons  originating in the same primary CR interaction. For the ANTARES detector the background contribution from muon events originating from independent  showers is negligible.

\noindent \textbf{Simulation of the detector.}
In the simulation of the  digitized signal, the main features of the PMTs and of the ARSs are taken into account. The simulated photons arriving on each PMT are used to determine the charge of the analogue pulse; the charge of consecutive pulses are added during the 25 ns integration time. 
The hit time is determined by the arrival time of the first photon.
To simulate the noise in the apparatus, background hits generated according to the distribution of a typical data run are  added. 
The status of each of the 885 OMs in this particular run is also reproduced. 
The OM simulation also includes the probability of a detected hit giving rise to an afterpulse in the PMT.
This probability was measured  in the laboratory \cite{anta_nim} and was confirmed with  deep-sea data.

\section{Event selection and background rejection} \label{real_detector}

The data were collected during the period from December 2007 to December 2009 with 9, 10 and 12 active line configurations. 
The  runs were selected according to a set of data-quality  criteria described in \cite{mar5line}; in particular a baseline rate $< 120$ kHz and a burst fraction $< 40\%$. 
A total of 3076 runs satisfy the conditions. The total live time is 334 days:  70 days with 12 lines, 128 days with 10 lines and 136 days with 9 lines.
In the detector simulation three different configurations are taken into account, based on the number of active lines.  
For each detector geometry, a typical run  is selected  to reproduce  on average the conditions and the background in the data.

\subsection{Rejection of atmospheric muons}\label{step1}
The ANTARES trigger rate, which is dominated by atmospheric muons, is a few Hz. The reconstruction algorithm \cite{aart_icrc09} results in approximately 5\% of triggered downgoing muons to be mis-reconstructed as upgoing. This contamination can be readily reduced by applying requirements on the geometry of the event and on the track reconstruction quality parameter $\Lambda$.
For simulated upgoing atmospheric neutrino events the $\Lambda$ distribution has a maximum around -4.5 and 95\% of  the events have $\Lambda>-5.5$. Two steps are used to remove the contamination of mis-reconstructed  atmospheric muons  from the final sample.

\noindent \textbf{First-level cuts.} Selection of 
$(i)$ upgoing particles with reconstructed zenith angle $\theta_{rec} < 80^\circ$ (corresponding to 0.83$\times 2\pi$ sr); 
$(ii)$ ${\Lambda > -6}$; $(iii)$ ${N_{hit} > 60}$;
$(iv)$ reconstruction with at least two lines. 
The first-level cuts remove all MC atmospheric muons with $E_T<$ 1 TeV and reduce the rate of mis-reconstructed events by almost 3 orders of magnitude, as indicated in Table \ref{tab:prelim}.
\begin{table}[tdp]
\centering
{\small \begin{tabular}{|c|c|c|c||c|}
\hline
      &  $\mu_{Atm}$ & $\nu_{Atm}$ & $\nu_{sig}$ & Data\\ 
\hline
\hline
{Reco} &  $2.2 \times10^8$ & $7.11 \times 10^3$ & $106$ & $2.5 \times10^8$  \\ \hline
{Upgoing} &$4.8 \times10^6$ & $5.50 \times 10^3$ & $80$ & $5.2 \times10^6$  \\ \hline
{1$^{st}$-level} & $9.1 \times10^3$ & $142$ & $24$ & $1.0 \times10^4$  \\ \hline
{2$^{nd}$-level} & 0 & 116 & 20 & $134$ \\
\hline
\end{tabular}
}
\caption{{\small Expected event number in 334 days of equivalent live time for the three MC samples (atmospheric muons, atmospheric neutrinos (Bartol+RQPM), astrophysical signal from Eq. \ref{test_limit} and data. Reco: at the reconstruction level; Upgoing:  reconstructed as upgoing; {1$^{st}$-level}: after the first-level  cuts; {2$^{nd}$-level}: after the second-level cut. The number of events in the data set is known only at the end, after the un-blinding procedure, see Sec. 4.}}
\label{tab:prelim} 
\end{table}

\noindent \textbf{Second-level cut.} 
The remaining mis-reconstructed atmospheric muons have a quality parameter $\Lambda$ which on average decreases with increasing $N_{hit}$.
Values of a cut parameter ($\Lambda^*$) are obtained in 10 different intervals of $N_{hit}$,   in order to reduce the expected rate of mis-reconstructed events to less than 0.1$\div$ 0.3 events/year in each  interval.
A parametrization of the values of $\Lambda^*$ as a function of $N_{hit}$ is: 
\begin{equation}
\Lambda^* = \left\{ \begin{array}{ll}-4.59-5.88 \cdot 10^{-3} N_{hit} \ \ & \textrm{for }\ N_{hit} \leq 172  \\
	 -5.60 &  \textrm{for }\ N_{hit} >   172    \end{array}\right  .
\label{eq:intercut}
\end{equation}
Removing all events with $\Lambda<\Lambda^*$,  the atmospheric muons are completely suppressed (last row of Table \ref{tab:prelim}). Independent MC atmospheric muon simulations using CORSIKA (see details in \cite{mar5line}) confirm that the maximum contamination in the final sample is less than 1 event/year. The effects of the first- and second-level cuts on signal and atmospheric neutrinos  are also given in Table \ref{tab:prelim}.

\subsection{The energy estimator}\label{ene_ext}
\begin{figure}[tbh]
{\centering 
\resizebox*{!}{0.40\textheight}{\includegraphics{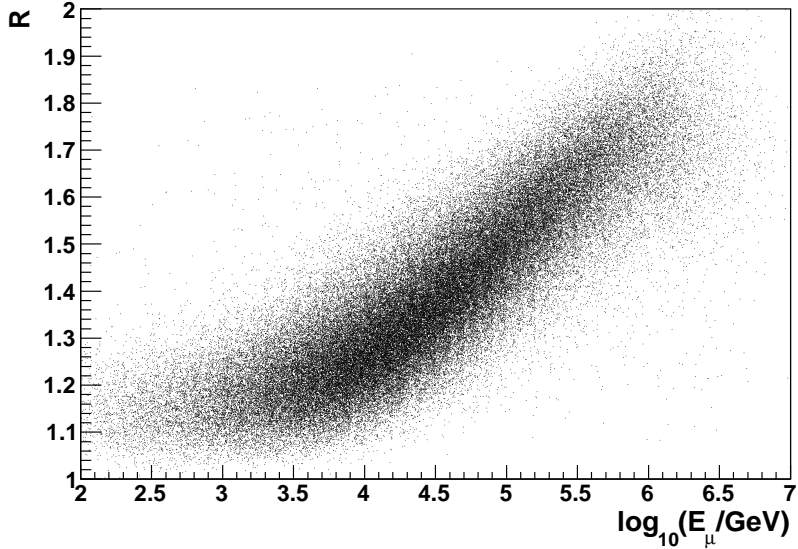}}\par}
\caption{{\small Mean number of repetitions as a function of the true neutrino-induced muon energy for events passing the second-level cut.}}
\label{R_vs_E}
\end{figure}
To separate atmospheric and astrophysical neutrinos,  an original energy estimator is defined, which is  based on hit repetitions  in the OMs due to the different arrival time of \textit{direct} and \textit{delayed} photons. 
 The number of repetitions $R_i$ for the $i$-th OM is defined as the number of hits in the same OM within 500 ns from the earliest hit selected by the reconstruction algorithm. In most cases, $R_i$ =1 or 2. The mean number of repetitions in the event is defined as $R = \frac{\sum R_i}{N_{OM}} $, where $N_{OM}$ is the number of OMs in which hits used by the tracking algorithm are present.
After the second-level cut, $R$ is linearly correlated with the log of the true muon energy $ E_{true}$ in the range from 10 TeV to 1 PeV, see Fig. \ref{R_vs_E}.
$R$ slightly saturates after 1 PeV. The distribution of log$(E_{rec}/E_{true})$ has a HWHM=0.4 when $R$ is used as an estimator of the muon energy  $E_{rec}$. This  energy estimator is robust because it does not depend on the number of active OMs and on non-linear effects on charge integration. 

\begin{figure}[tbh]
{\centering 
\resizebox*{!}{0.40\textheight}{\includegraphics{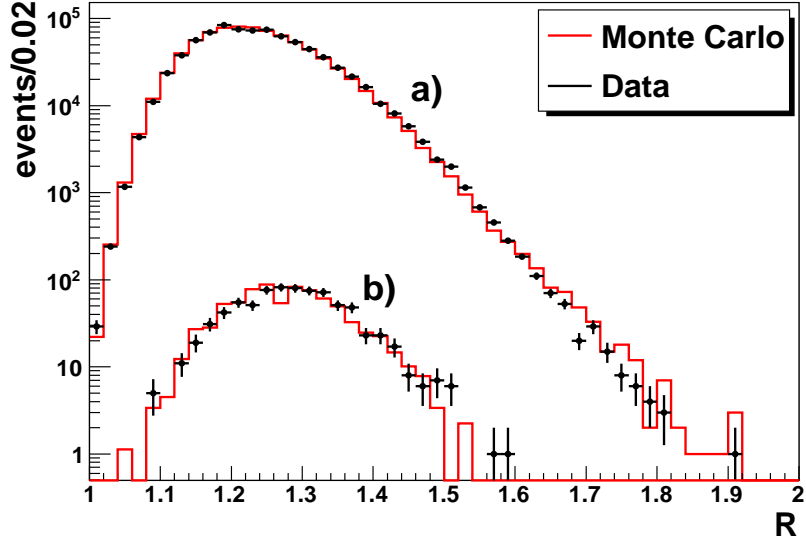}}\par}
\caption{{\small Distributions of the number of events as a function of the $R$ energy estimator for data and MC before the second-level cut. a) downgoing muons; b) mis-reconstructed as upgoing muons. The points are 12 line data, the histograms show the atmospheric muon MC normalized to the data.}}
\label{down_prelim}
\end{figure}

Atmospheric muons are used to check the agreement between data and MC for the $R$ variable. 
The first-level  cuts (Sec. \ref{step1}) are applied both to data and MC, except that tracks reconstructed as downgoing $(\theta_{rec} > 90^\circ)$ are selected. In the data set, $1.37\times 10^7$ events are present, and $1.22\times 10^7$ in the simulation for the corresponding live time. 
The distribution of the $R$ variable is shown in Fig. \ref{down_prelim} a) for a subset of 20 days live time of the 12 line data. 
A second comparison in Fig. \ref{down_prelim} b) uses those atmospheric muons that survived the first-level cuts in the same data set and are mis-reconstructed as upgoing (the true upgoing atmospheric neutrinos are about 1.5\% of the total, see Table \ref{tab:prelim}). The MC curve a) was normalized to the data by a factor 1.12, and b) by a factor 1.15. These factors are well within the overall systematic uncertainties on the atmospheric muon flux, and the relative difference is accounted for by the uncertainty on the OMs angular acceptance \cite{mar5line}. 
\begin{figure}
{\centering 
\resizebox*{!}{0.40\textheight}{\includegraphics{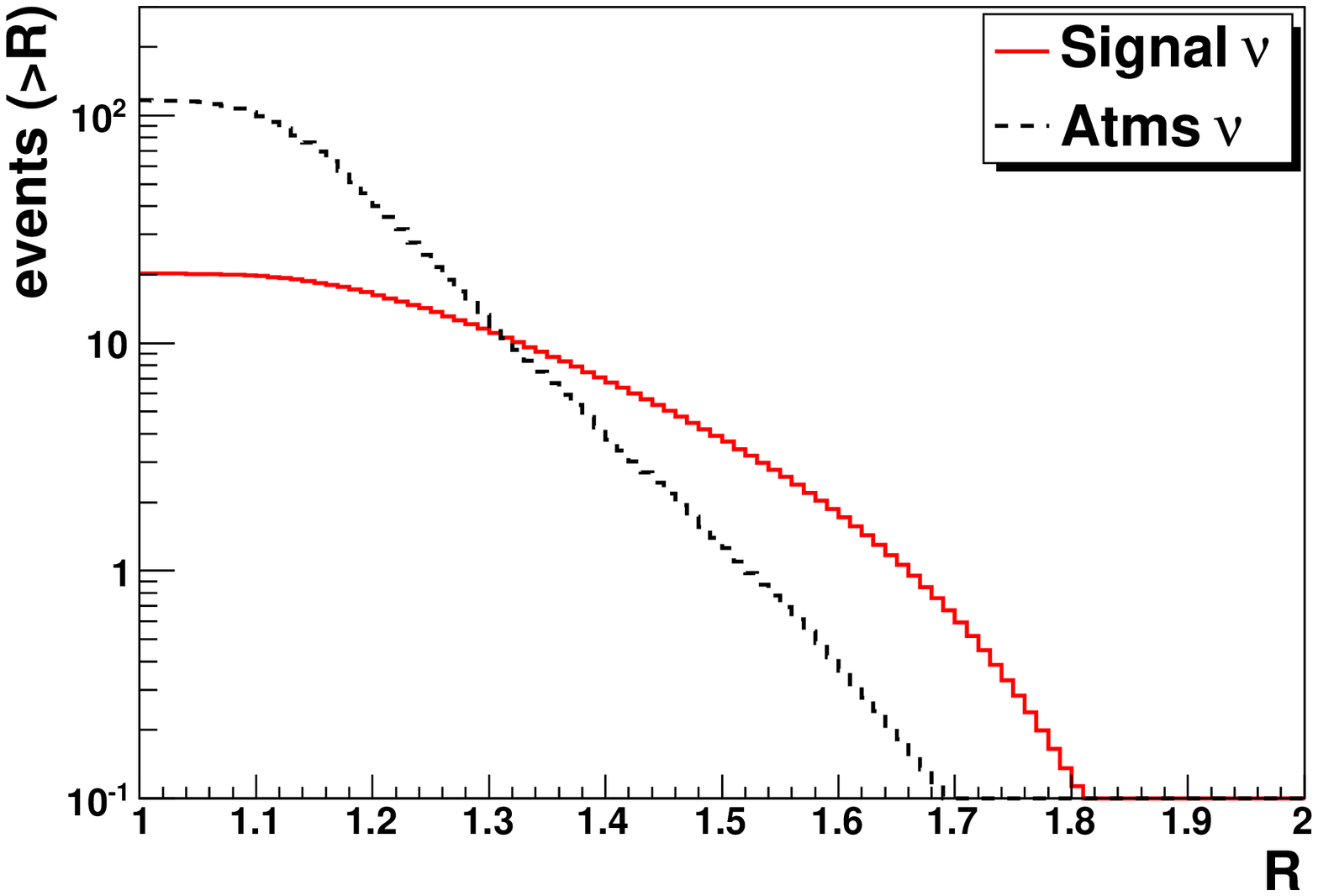}}\par}
\caption{{\small Cumulative distributions of the $R$ variable  for simulated diffuse flux signal (Eq. \ref{test_limit}) and atmospheric neutrino events (including the prompt from the RQPM model). 
}}
\label{fig:MRF_all}
\end{figure}

\subsection{Signal/atmospheric $\nu_\mu$  background discrimination}\label{nu_atm_discr}
The separation of the diffuse flux signal from the atmospheric $\nu_\mu$ background is performed by a cut on the $R$ variable. In order to avoid any bias, a blinding procedure on MC events is applied, without using information from the data.  
The numbers of expected events for signal ($n_s$) and  background ($n_b$) are computed as a function of $R$ to find the optimal cut value of $R$. 
  Later the number of observed data events ($n_{obs}$) are revealed (\textit{un-blinding procedure}) and compared with the expected background  for the selected region of $R$. If this number is compatible with the background, the upper limit  for the flux at a 90\% confidence level (c.l.) is calculated using the Feldman-Cousins method \cite{feldman}.

Simulated atmospheric neutrino events are used also   to calculate the ``average upper limit'' that would be observed by an ensemble of hypothetical experiments with no true signal ($n_s =0$) and expected background $n_b$.   Taking into account all the possible fluctuations for the estimated background, weighted according to their Poisson probability of occurrence, the average upper limit is:
\begin{equation}
\overline\mu_{90\%}(n_b) = \sum_{n_{obs}=0}^\infty \mu_{90\%}(n_{obs},n_b) \frac{(n_b)^{n_{obs}}}{(n_{obs})!} e^{-n_b}  .
\label{eq:aul}
\end{equation}
The best average upper limit is obtained with the cut on the energy estimator that minimizes the so-called Model Rejection Factor \cite{mrp}, 
$\textrm{MRF} = \frac{\overline\mu_{90\%}(n_b)}{n_s}$, 
and hence minimizes the average flux upper limit:
\begin{equation}
\overline \Phi _{90\%} = \Phi_\nu   \cdot \frac{\overline\mu_{90\%}(n_b)}{n_s} = \Phi_\nu \cdot \textrm{MRF}  .
\label{eq:sensitivity}
\end{equation} 

The value of $R$ which minimizes the MRF function  in Eq. \ref{eq:sensitivity} is used as the discriminator between \textit{low energy} events, dominated by the atmospheric neutrinos, and \textit{high energy} events, where the signal could exceed  the background. 

\begin{figure}
{\centering 
\resizebox*{!}{0.40\textheight}{\includegraphics{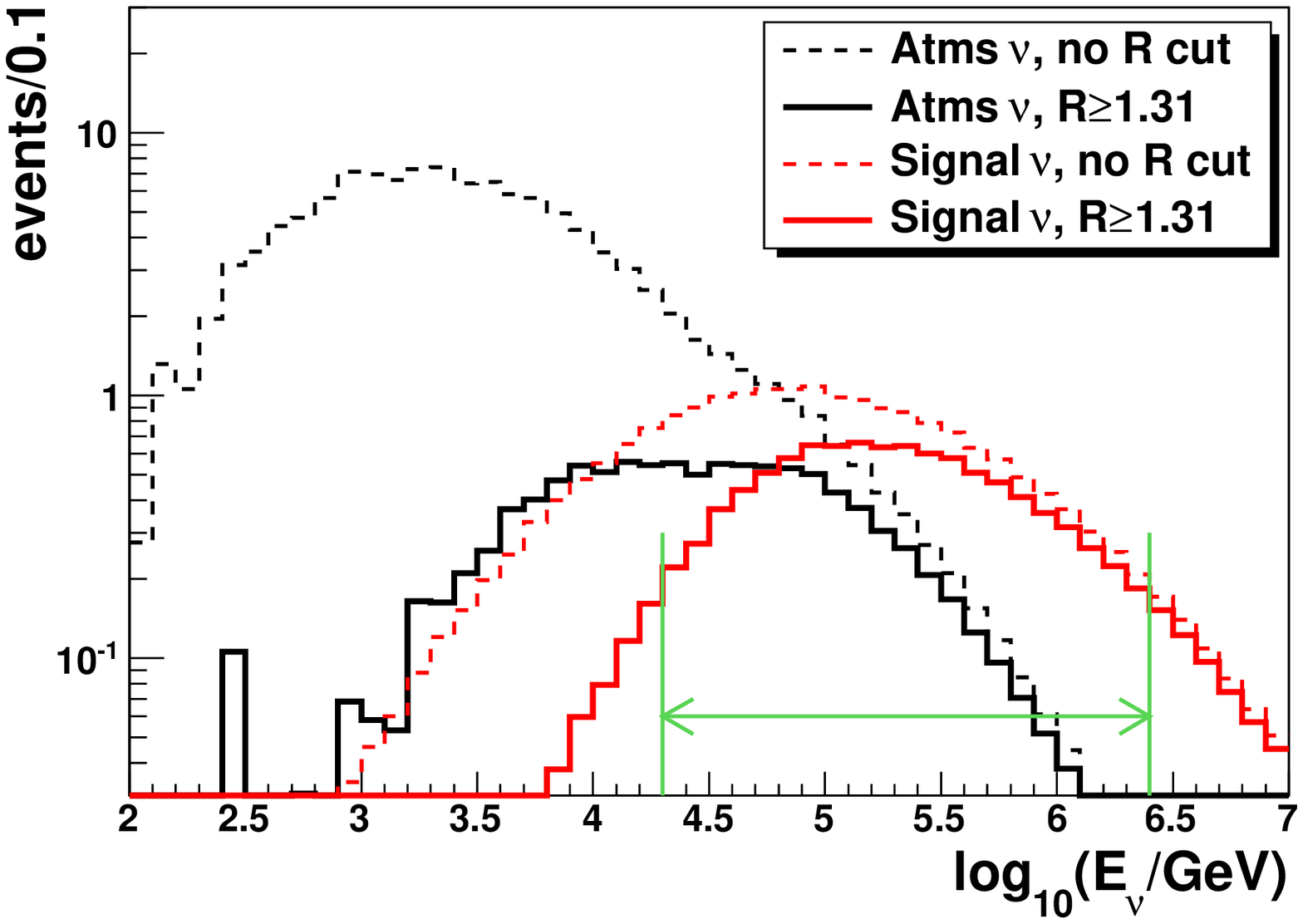}}\par}
\caption{{\small Signal and background neutrino energy spectra as a function of the true neutrino energy after the second-level cut with and without the requirement $R\ge 1.31$.  The energy region containing 90\% of the astrophysical neutrino signal  is indicated. }}
\label{energia_all}
\end{figure}

The method relies on knowledge of the number of background events expected for a given period of data. 
The cumulative distributions of the $R$ variable are computed for atmospheric neutrino background and diffuse flux signal for the three discussed  configurations of the ANTARES detector and the corresponding live times.
For the atmospheric neutrino background, the conventional flux and the prompt models are considered separately. 
Fig. \ref{fig:MRF_all} shows the cumulative distributions of the $R$ variable for signal and background neutrinos (Bartol+RQPM).
Using these cumulative distributions, the  MRF  is calculated as a function of  $R$; the minimum (MRF=0.65) is found for $R=1.31$. 
Assuming  the Bartol (Bartol+RQPM) atmospheric $\nu_\mu$ fluxes, 8.7 (10.7) background events and  10.8 signal events (assuming the test flux of Eq. \ref{test_limit}) are expected for $R\ge 1.31$.
Fig. \ref{energia_all}  shows the energy spectra for signal and background neutrino events before and after the cut   $R \ge 1.31$. The central 90\% of the signal is found  in the neutrino energy range   $20 \  \mathrm{TeV} < E_{\nu} < 2.5 \  \mathrm{PeV}$.   

\begin{table}[htb]
\centering{\small
\begin{tabular}{|c|c|c||c|c|}
\hline
Flux model & N$_{< 1.31}$	& D/ N$_{< 1.31}$ & N$_{\ge 1.31}$ & N$^*_{\ge 1.31}$ \\
\hline \hline
Bartol	& 104.0   & 1.20 & 8.7 & 10.4 \\ 
\hline
Bartol+ RQPM & 105.2 & 1.19 & 10.7 & 12.7 \\ \hline
\end{tabular}
\caption{{\small Number of expected atmospheric neutrino events in the two intervals of $R$. Atmospheric $\nu_\mu$  according to the Bartol flux and the Bartol flux plus the prompt contribution from the RQPM model are considered separately. 
N$_{< 1.31}$ is for $R <$ 1.31, where D=125 data events are observed.
N$_{\ge 1.31}$ (N$^*_{\ge 1.31}$) is the number of expected
background events for $R\ge 1.31$, without (with) the normalization by the factor D/N$_{< 1.31}$. }} \label{tab:events} }
\end{table}

\section{Data un-blinding and results}\label{data}

\begin{figure}[tb!]
{\centering 
\resizebox*{!}{0.40\textheight}{\hskip -0.5cm\includegraphics{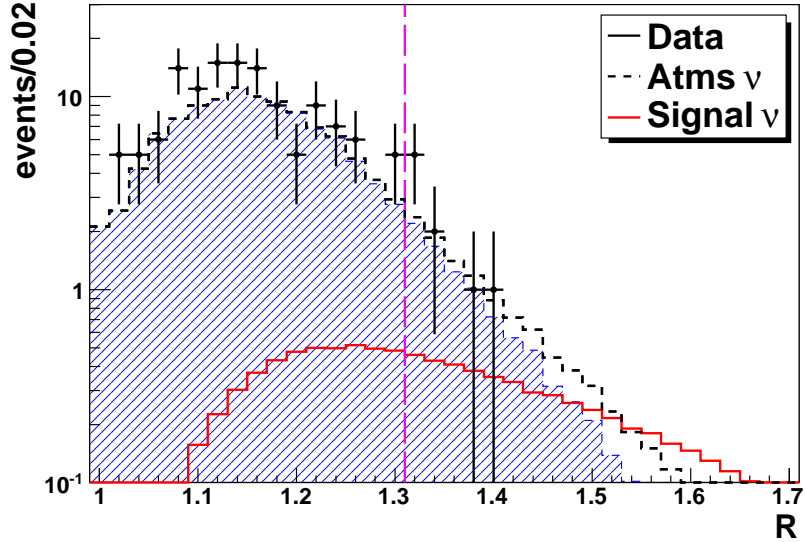}}\par}
\caption{{\small Distribution of the $R$ parameter for the 134 neutrino candidates in the 334 days of equivalent live time. Points represent data, the filled histogram is the atmospheric neutrino MC (Bartol model only). The dashed line represents the maximum contribution (RQPM) of  ``prompt'' neutrinos. The MC predictions are not normalized to the data. The signal at the level of the upper limit (Eq. \ref{eq:upper_limit}) is shown as a full line. The cut at $R=1.31$ is indicated as a vertical line.}}
\label{fig:Neutrino_unblinded}
\end{figure}

Events surviving the second-level cut are upgoing neutrino candidates.
Fig.  \ref{fig:Neutrino_unblinded} shows the  distribution  of the  neutrino candidates as a function of $R$,  compared with  that given by the  atmospheric neutrino MC. At this stage, only the 125 events with $R<1.31$ are un-blinded. The events with $R\ge 1.31$ in Fig. \ref{fig:Neutrino_unblinded} are revealed only after the un-blinding of the data samples.
The number of expected events is lower by $\sim$ 20\% with respect to the detected events (D). 
This discrepancy is well within the systematic uncertainties of the absolute neutrino flux at these energies (25-30\%) \cite{bartol}.

Table  \ref{tab:events} shows the number of expected MC events N$_{\ge 1.31}$ and N$^*_{\ge 1.31}\equiv \textrm{N}_{\ge 1.31}\cdot \textrm{D/N}_{< 1.31}$ both for the conventional Bartol and Bartol+RQPM fluxes.
Most prompt models give negligible contribution (the average over all considered models gives 0.3 events), the RQPM model predicts the largest contribution of 2.0 additional events with respect to the conventional Bartol flux.
After data/MC normalization in the $R<1.31$ region, the number of expected background events for $R\ge 1.31$ from a combined model of Bartol flux plus the average contribution from prompt models 
is $10.7$ events.

A reasonable agreement between data and MC for the $R$ distribution both for atmospheric muons (c.f. Fig. \ref{down_prelim}) and for atmospheric neutrinos in the test region $R<$1.31 (c.f. Fig. \ref{fig:Neutrino_unblinded})  is found. Consequently the data was un-blinded for the signal region $R\ge$ 1.31 and 9 high-energy neutrino candidates are found. 

Systematic uncertainties on the expected number of background events in the high energy region ($R\ge 1.31$) include: 
$(i)$ the contribution of prompt neutrinos, estimated as ${}^{+1.7}_{-0.3}$ events. In the following, the largest value is conservatively used. 
$(ii)$ The uncertainties from the neutrino flux from charged meson decay as a function of the energy. 
By changing the atmospheric neutrino spectral index by $\pm 0.1$, both below and above $\sim$10 TeV (when the conventional neutrino flux has spectral index one power steeper than that of the primary CR below and after the knee, respectively), the relative number of events for $R\ge 1.31$ changes at most by $\pm 1.1$, keeping in the region $R<1.31$ the number of MC events equal to the number of data. The migration from the Bartol to the Honda MC \cite{honda} produces a smaller effect. 
The uncertainties on the detector efficiency (including the angular acceptance of the optical module \cite{mar5line}, water absorption and scattering length, trigger simulation and the effect of   PMT afterpulses) amount to 5\% after the normalization to the observed atmospheric $\nu_\mu$ background in the test region.  

\begin{figure}[tb!]
{\centering 
\resizebox*{!}{0.5\textheight}{\hskip -1.1cm \includegraphics{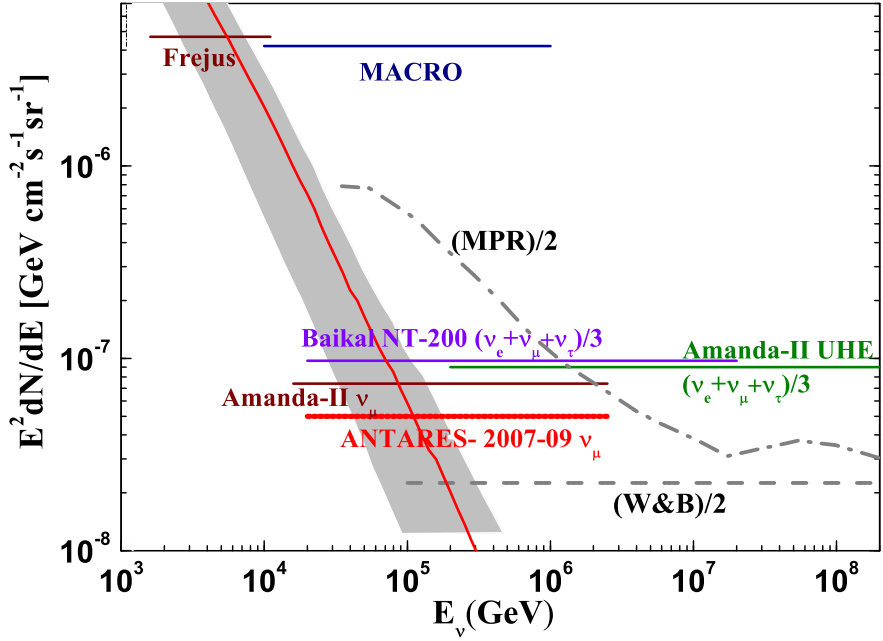}}\par}
\vskip -0.6cm \caption{The ANTARES 90\% c.l. upper limit for a $E^{-2}$ diffuse high energy $\nu_\mu+\overline \nu_\mu$ flux obtained in this work, compared with the limits  from  other experiments. 
The Frejus \cite{Frejus}, MACRO \cite{MACRO}, Amanda-II  2000-03 \cite{Amanda_numu} limits refer to $\nu_\mu+\overline \nu_\mu$.
The Baikal \cite{Baikal} and Amanda-II  UHE 2000-02 \cite{Amanda-UHE} refer to neutrinos and antineutrinos of all-flavours, and are divided by 3. For reference, the W\&B \cite{wb} and the MPR\cite{mpr} upper bounds for transparent sources are also shown. They are divided by two, to take into account neutrino oscillations. 
The grey band represents the expected variation of the atmospheric $\nu_\mu$ flux: the minimum is the Bartol flux from the vertical direction; the maximum the Bartol+RQPM flux from the horizontal direction. The central line is averaged over all directions.}
\label{fig:upper_limits}
\end{figure}

The number of observed events is compatible with the number of expected background events. The $90\%$  c.l. upper limit on the number of signal events $\mu_{90\%}(n_b)$ for $n_b =10.7\pm 2$ background events and $n_{obs}=9$ observed events including the systematic uncertainties is computed with the method of \cite{conra}. 
The value $\mu_{90\%}(n_b) = 5.7$ is obtained. The profile likelihood method \cite{rolke} gives similar results.
The corresponding flux upper limit is given by $\Phi_{90\%} = \Phi_\nu \cdot \mu_{90\%}/ n_s$:
\begin{equation}
E^2 \Phi _{90\%}  =   5.3 \times 10^{-8}   \  \mathrm{GeV\ cm^{-2}\ s^{-1}\ sr^{-1}} 
\label{eq:upper_limit}
\end{equation}
(our expected sensitivity is $7.0 \times 10^{-8}$ GeV cm$^{-2}$  s$^{-1}$  sr$^{-1}$). 
This limit holds for the energy range
between  20 TeV to 2.5 PeV, as shown in Fig. \ref{energia_all}.  The result is compared with other measured flux upper limits in Fig. \ref{fig:upper_limits}\footnote{Charged current $\nu_\tau$ interaction can contribute via $\tau^- \rightarrow \mu^- \nu_\tau \overline \nu_\mu$ (and similarly the  $\overline \nu_\tau$) by less than $\sim$ 10\% both for signal and background. For the background, the $\nu_\tau$ contribution is almost completely absorbed by the uncertainty on the overall normalization, while it is neglected in the signal.}.

A number of models predict cosmic neutrino fluxes with a spectral shape different from $E^{-2}$.  
For each model a cut value $R^*$ is optimized following the procedure in Sec. 3.3. Table \ref{tab:models} gives the  results for the models tested; the value of  $R^*$; the number $N_{mod}$ of $\nu_\mu$ signal events for $R\ge R^*$; the energy interval where 90\% of the signal is expected; the ratio between $\mu_{90\%}$ (computed according to \cite{feldman}) and $N_{mod}$. A value of  $\mu_{90\%}/N_{mod}<1$ indicates that the theoretical model is inconsistent with the experimental result at the 90\% c.l. In all cases (except for \cite{m95}), our results improve upon those obtained in \cite{Amanda_numu,Baikal,Amanda-UHE}. 

\section{Conclusions}
A  search for a diffuse flux of  high energy muon neutrinos from astrophysical sources  with the data from 334 days of live time of the ANTARES neutrino telescope is presented. A robust energy estimator, based on the mean number $R$ of repetitions of hits on the same OM  produced by  direct and delayed photons in the detected muon-neutrino events, is used. 
The $90\%$  c.l. upper limit for  a $E^{-2}$ energy spectrum is $E^2 \Phi _{90\%}  =   5.3 \times 10^{-8}   \  \mathrm{GeV\ cm^{-2}\ s^{-1}\ sr^{-1}}$ in the energy range  20 TeV -- 2.5 PeV.  
Other models predicting cosmic neutrino fluxes with a spectral shape different from $E^{-2}$ are tested and some of them excluded at a 90\% c.l..

\begin{table}[tdp]
\centering{\small
\begin{tabular}{|c|c|c|c|c|}
\hline
 Model   & R$^*$ & N$_{mod}$ & $\Delta E_{90\%}$ & $\mu_{90\%}/N_{mod}$ \\ 
& & &(PeV)  &\\
\hline
\hline
MPR\cite{mpr}  & 1.43 & 3.0 & 0.1$\div$ 10 & 0.4 \\ \hline
P96$p\gamma$\cite{p96}  & 1.43 & 6.0 & 0.2$\div$ 10 & 0.2 \\ \hline
S05\cite{s05}  & 1.45 & 1.3 & 0.3$\div \ $ 5 & 1.2 \\ \hline
SeSi\cite{sesi}& 1.48 & 2.7 & 0.3$\div$ 20 & 0.6 \\ \hline
M$pp+p\gamma$\cite{m95}  & 1.48 & 0.24 &0.8$\div$ 50  & 6.8 \\ \hline
\end{tabular}
}
\caption{{\small Astrophysical flux models, the value of the R$^*$ which minimizes the MRF, the expected number of events $N_{mod}$, the energy range $\Delta E_{90\%}$ in which the 90\% of events are expected, and the ratio $\mu_{90\%}/N_{mod}$. }}
\label{tab:models} 
\end{table}

\vskip 0.5cm
\noindent \textbf{Acknowledgements}
The authors acknowledge the financial support of the funding agencies: Centre National de la Recherche Scientifique (CNRS), Commissariat \'a l'\'energie atomique et aux energies alternatives  (CEA), Agence National de la Recherche (ANR), Commission Europ\'enne (FEDER fund and Marie Curie Program), R\'egion Alsace (contrat CPER), R\'egion Provence-Alpes-C\^ote d'Azur, D\'epartement du Var and Ville de La Seyne-sur-Mer, France; Bundesministerium f\"ur Bildung und Forschung (BMBF), Germany; Istituto Nazionale di Fisica Nucleare (INFN), Italy; Stichting voor Fundamenteel Onderzoek der Materie (FOM), Nederlandse organisatie voor Wetenschappelijk Onderzoek (NWO), The Netherlands; Council of the President of the Russian Federation for young scientists and leading scientific schools supporting grants, Russia; National Authority for Scientific Research (ANCS), Romania; Ministerio de Ciencia e Innovaci\'on (MICINN), Prometeo of Generalitat Valenciana (GVA) and MultiDark, Spain. We also acknowledge the technical support of Ifremer, AIM and Foselev Marine for the sea operation and the CC-IN2P3 for the computing facilities.









\end{document}